%% file: wellnotsymm7.tex
\documentclass[11pt]{article}

\usepackage{amsmath}
\usepackage{amssymb}
\usepackage{latexsym}
\usepackage{graphicx}
\usepackage{epsfig}
\usepackage{stmaryrd}
\usepackage{framed}
\usepackage{bm}
\usepackage{tikz}

\newtheorem{thm}{\bf Theorem}[section]
\newtheorem{prop}[thm]{\bf Proposition}  
\newtheorem{cor}[thm]{\bf Corollary}
\newtheorem{deff}[thm]{\bf Definition}

\newtheorem{remark}[thm]{\bf Remark}

\def\Bbox{
{\unskip\nobreak\hfil\penalty50
\hskip1em\hbox{}\nobreak\hfil{\lower .5pt \hbox{$\Box$}}
\parfillskip=0pt \finalhyphendemerits=0 \par}
}

\def\eop{
\ifmmode {\hbox{\Bbox}} \else \Bbox \fi
}

\def\bbox{
\ifmmode {\hbox{\bbox}} \else \Bbox \fi
}

\newcommand{\circarrow}{\stackrel{\bullet}{\rightarrow}}
\newcommand{\CL}{\mathrm{CL}}
\newcommand{\Set}{\mathrm{SET}}
\newcommand{\bCL}{\mathbf{CL}}

\newcommand{\x}{\times}
\newcommand{\op}{{\rm op}}
\newcommand{\id}{\mathrm{id}}
\newcommand{\bid}{\mathbf{id}}
\newcommand{\bpi}{{\bm{\pi}}}
\newcommand{\four}{{\mathbf{4}}}

\newcommand{\bDelta}{\bm{\Delta}}
\newcommand{\brho}{\bm{\rho}}
\newcommand{\blangle}{\bm{\langle}{\kern -.5em }\bm{\langle}}
\newcommand{\brangle}{\bm{\rangle}{\kern -.5em }\bm{\rangle}}
\newcommand{\two}{\mathbf{2}}

\newcommand{\sCL}{\mathbf{SCL}}
\newcommand{\ACL}{\mathbf{ACL}}
\newcommand{\AsCL}{\mathbf{EACL}}
\newcommand{\CCL}{\mathbf{CCL}}

\begin{document}

\title{{\bf An analysis of the equational properties of the well-founded fixed point}}

\author{
Arnaud Carayol\\
Universit\'e Paris Est and CNRS\\
France\\
\and 
Zolt\'an \'Esik\thanks{The second author received support from NKFI grant no. ANN 110883
and the Universit\'e Paris Est.} 
\\
University of Szeged\\
Hungary
}

\maketitle

\begin{abstract}
Well-founded fixed points have been used 
in several areas of knowledge representation 
and reasoning and to give semantics 
to logic programs involving negation. 
They are an important ingredient of approximation
fixed point theory. We study the logical properties of the 
(parametric) well-founded fixed point operation. We show that 
the operation satisfies several, but  not all of the 
equational properties of fixed point operations 
described by the axioms of iteration theories. 
\end{abstract} 

\section{Introduction} 

Fixed points and fixed point operations have been used in just about all areas of 
computer science. There has been a tremendous amount of work on the existence, 
construction and logic of fixed point operations. 
It has been shown that most fixed point operations, including 
the least (or greatest) fixed point operation on monotonic functions 
over complete lattices, satisfy the same equational properties.
These equational properties are  captured by the notion of iteration 
theories, or iteration categories, cf. \cite{BEbook}
or the recent survey \cite{EsMFCS2015}.

For an account of fixed point approaches to logic programming containing 
original references   we refer to \cite{Fittingsurvey}. These approaches, 
and in particular 
the stable and well-founded fixed point semantics of logic
programs with negation, based on the notion of 
bilattices, have led to the development of an 
elegant abstract `approximation fixed point theory', cf.  
\cite{Deneckeretalsurvey,DeneckeretalULT,Vennekensetal}.

In this paper, we study the equational properties of the well-founded fixed 
point operation as defined in \cite{Deneckeretalsurvey,DeneckeretalULT,Vennekensetal}
with the aim of relating well-founded fixed points to iteration categories. 
We extend the well-founded fixed point operation to a parametric operation
giving rise to an external fixed point (or dagger) operation \cite{BEbook,BEccc} over 
the cartesian category of approximation function pairs between complete 
bilattices. We offer 
an initial analysis of the equational properties of the well-founded 
fixed point operation. Our main results show that several identities of iteration theories 
hold for the well-founded fixed point operation, but some others fail.

\section{Complete lattices and bilattices}  

Recall that a \emph{complete lattice} \cite{Daveyetal} is a partially ordered set $L = (L,\leq)$ such that
each $X \subseteq L$ has a supremum $\bigvee X$ and hence also an infimum
$\bigwedge X$. In particular, each complete lattice has a least 
and a greatest element, respectively denoted either $\bot$ and $\top$, 
or $0$ and $1$.  
We say that a function $f: L \to L$ over a complete lattice $L$
is monotonic (anti-monotonic, resp.) if for all $x,y \in L$, if $x \leq y$ 
then $f(x) \leq f(y)$ ($f(x) \geq f(y)$, resp.). 

A \emph{complete bilattice}\footnote{Sometimes bilattices are equipped with a negation operation
and the bilattices as defined here are called pre-bilattices.}  
\cite{Fittingsurvey,Fittingnice,Ginsberg} $(B,\leq_p,\leq_t)$ is equipped with two partial orders, 
$\leq_p$ and $\leq_t$, both giving rise to a complete lattice. 
We will denote the $\leq_p$-least and greatest elements of a complete bilattice 
by $\bot$ and $\top$, and the $\leq_t$-least and greatest elements by $0$ and $1$,
respectively.  

An example, depicted in Figure~\ref{figure:four}, of a complete bilattice is $\four$, which has 4 elements, $\bot,\top,0,1$. 
The nontrivial order relations are given by $\bot \leq_p 0,1 \leq_p \top$ and $0 \leq_t \bot,\top \leq_t 1$. 

\input{four.tex}

Two closely related  constructions of a complete bilattice from a complete lattice
are described in \cite{Deneckeretalsurvey} and \cite{Fittingnice}, see \cite{Ginsberg}
for the origins of the constructions. 
Here we recall one of them. Suppose that $L = (L,\leq)$ is a complete lattice
with extremal (i.e., least and greatest) elements $0$ and $1$. 
Then define the partial orders $\leq_p$ and $\leq_t$ on $L \x L$ 
as follows: 
\begin{eqnarray*}
(x,x') \leq_p (y,y') &\Leftrightarrow& x \leq y\ \wedge \ x' \geq y'\\
(x,x') \leq_t (y,y') &\Leftrightarrow& x \leq y \ \wedge\ x' \leq y'.
\end{eqnarray*} 
Then $L \x L$ is a complete bilattice with $\leq_p$-extremal elements 
$\bot = (0,1)$ and $\top = (1,0)$, and $\leq_t$-extremal elements $0 = (0,0)$ and $1 = (1,1)$. 
Note that when $L$ is the $2$-element lattice $\mathbf{2} = \{0 \leq 1\}$, then 
 $L \x L$ is isomorphic to $\four$.
In this paper, we will mainly be concerned with the 
ordering $\leq_p$. 

In any category, we usually denote the composition of morphisms $f: A \to B$ 
and $g: B \to C$ by $g \circ f$ and the 
identity morphisms by $\id_A$. 
We let $\Set$ denote the category of sets and functions
and we denote by  $\CL$ the category of complete lattices and monotonic functions.
Both $\Set$ and $\CL$ have all products and hence are 
\emph{cartesian categories}. The 
usual direct product, equipped with the pointwise order in $\CL$, 
serves as categorical product. In $\CL$, a terminal object is a $1$-element lattice 
$T$. In both categories, 
for any sequence $A_1,\ldots,A_n$ of objects, the categorical projection morphisms 
$\pi^{A_1 \x \cdots \x A_n}_i: A_1 \x\cdots \x A_n  \to A_i$, $ i \in [n] = \{1,\ldots ,n\}$, are 
the usual projection functions. 

Products give rise to a \emph{tupling} operation.  Suppose that $f_i:  C \to A_i$, $i \in [n]$ in $\Set$ or $\CL$, 
or in any cartesian category. Then there is a unique
$f: C \to A_1 \x \cdots \x A_n$ with $\pi^{A_1 \x \cdots \x A_n}_i \circ f = f_i$ 
for all $i \in [n]$. We denote this unique morphism $f$ by $\langle f_1,\ldots,f_n \rangle$ 
and call it the (target) tupling of the $f_i$ (or pairing, when $n = 2$). 

And when $f: C \to A$ and $g: D \to B$, then we define $f \x g$ 
as the unique morphism $h: C \x D \to A \x B$ with $\pi^{A \x B}_1 \circ h = 
f \circ \pi^{C \x D}_1$ and $\pi^{A \x B}_2 \circ h = g \circ \pi^{C \x D}_2$.

When $m,n \geq 0$, $\rho$ is a function $[m] \to [n]$ and 
$A_1,\ldots,A_n$ is a sequence of objects in a cartesian category, 
we associate with $\rho$ (and $A_1,\ldots,A_n$) the morphism
$$\rho^{A_1,\ldots,A_n} = \langle \pi^{A_1 \x \cdots \x A_n}_{\rho(1)}, \ldots,  \pi^{A_1\x \cdots \x A_n}_{\rho(m)} \rangle$$ 
from $A_1 \x \cdots \x A_n$ to $A_{\rho(1)} \x \cdots \x A_{\rho(m)}$ 
(Note that in $\Set$ and $\CL$, $\rho^{A_1,\ldots,A_n}$ maps $(x_1,\ldots,x_n) \in A_1 \x \cdots \x A_n$
to $(x_{\rho(1)},\ldots,x_{\rho(m)}) \in A_{\rho(1)} \x \cdots \x A_{\rho(m)}$.)
With a slight abuse of notation, we usually let $\rho$ denote this morphism
as well. Morphisms of this form are sometimes called \emph{base morphisms}. 
When $m = n$ and $\rho$ is a bijection, then the associated morphism 
$ A_1\x \cdots \x A_n  \to A_{\rho(1)}\x \cdots \x A_{\rho(n)}$ 
is an isomorphism.  Its inverse is the morphism associated 
with the inverse $\rho^{-1}$ of the function $\rho$. 
For each object $A$,  
the base morphism associated 
with the unique function $[m] \to [1]$ is the 
\emph{diagonal morphism} $\Delta^A_m = \langle \id_A,\ldots,\id_A \rangle : A \to A^m$, usually denoted just $\Delta_m$.

\section{Iteration categories} 

The category $\CL$ is equipped with an (external) \emph{fixed point} or 
\emph{dagger} operation \cite{BEbook,BEccc} 
mapping a monotonic function $f: A \x B \to A$ to the monotonic function $f^\dagger : B \to A$ 
such that for all $y \in B$, $f^\dagger(y)$ is the least solution of the fixed 
point equation $x = f(x,y)$. We will sometimes denote $f^\dagger(y)$ by $\mu x. f(x,y)$.
It provides the unique least solution to the parametric fixed point equation 
\begin{eqnarray}
\label{eq-fp}
x &=& f(x,y).
\end{eqnarray}
When $B$ is the terminal object $T$, $f$ can be viewed as a function 
$A \to A$ and $f^\dagger$ can be identified with an element of $A$.

The least fixed point operation $^\dagger$ over $\CL$ satisfies several nontrivial identities
captured by the notion of \emph{iteration theories} or \emph{iteration categories} 
\cite{BEbook,EsMFCS2015}. 
For later use, we collect here some of these identities.

{\sc Fixed point identity}
\begin{eqnarray*}
f^\dagger &=& f \circ \langle f^\dagger, \id_B\rangle,
\end{eqnarray*}
where $f: A \x B \to A$.

The fixed point identity expresses that $f^\dagger(y)$ is a solution of the fixed point equation 
(\ref{eq-fp}). 

{\sc Parameter identity}
\begin{eqnarray*}
(f \circ (\id_A \x g))^\dag &=& f^\dag \circ g,
\end{eqnarray*}
for all $f: A \x B \rightarrow A$ and $g: C \rightarrow B$. 

In functional notation, the parameter identity expresses that 
if $h(x,z) = f(x,g(z))$, then for the least solution $h^\dagger(z)$ of the 
equation $x = h(x,z)$ it holds that $h^\dagger(z) = f^\dagger(g(z))$, where $f^\dagger(y)$ is the least 
solution of $x = f(x,y)$. 

{\sc Permutation identity}
\begin{eqnarray*}
(\rho \circ f \circ (\rho^{-1} \x \id_B))^\dagger &=&  \rho \circ f^\dagger,
\end{eqnarray*}
for all $f : A_1\x \cdots \x A_n \x B \rightarrow A_1\x \cdots \x A_n$
and permutation $\rho: [n]\to [n]$.

This can be explained alternatively as follows. Consider the (systems of) fixed point 
equations 
\begin{eqnarray}
\label{eq-fp2}
x &=& f(x,z)
\end{eqnarray} 
and 
\begin{eqnarray}
\label{eq-fp3}
y &=& \rho(f(\rho^{-1}(y),z)),
\end{eqnarray}
where $x$ ranges over $A_1 \x \cdots \x A_n$, $y$ ranges over $A_{\rho(1)}\x \cdots \x A_{\rho(n)}$ 
and $z \in B$. Here, $\rho$ also denotes the bijective function $A_1\x \cdots \x A_n \to A_{\rho(1)}\x \cdots \x A_{\rho(n)}$
as explained above, and $\rho^{-1}$ also denotes the inverse of this function. 
Then the permutation identity expresses that the least solution of (\ref{eq-fp3}) is $\rho(f^\dagger(z))$, 
where $f^\dagger(z)$ is the least solution of (\ref{eq-fp2}).

{\sc Composition identity}
\begin{eqnarray*}
(f \circ \langle g, \pi^{A \x C}_2\rangle)^\dagger 
&=& f \circ \langle (g \circ \langle f, \pi^{B \x C}_2\rangle)^\dagger, \id_C\rangle,
\end{eqnarray*}
where $f: B \x C \to A$ and $g: A \x C \to B$.  

The composition identity relates the fixed point equations
\begin{eqnarray}
\label{eq-eqcomp1}
x &=& f(g(x,z),z)
\end{eqnarray}
and
\begin{eqnarray}
\label{eq-eqcomp2}
y &=& g(f(y,z),z).
\end{eqnarray}
It asserts that the least solution of (\ref{eq-eqcomp1}) can be obtained 
by applying $f$ to the least solution of (\ref{eq-eqcomp2}) and the parameter. 

{\sc Double dagger identity}
\begin{eqnarray*}
f^{\dagger\dagger} &=& (f \circ (\langle \id_A,\id_A\rangle \x \id_B))^\dagger,
\end{eqnarray*}
for all $f: A \x A \x B \to A$.

This identity means that the least solution of the equation
\begin{eqnarray*}
x &=& f(x,x,z)
\end{eqnarray*}
is the same as the least solution of 
\begin{eqnarray*}
y &=& f^\dagger(y,z),
\end{eqnarray*}
 where $f^\dagger(y,z)$ is the least solution of $x = f(x,y,z)$.

{\sc Pairing identity}
\begin{eqnarray*}
\langle f,g\rangle^\dagger &=& 
\langle f^\dagger \circ \langle h^\dagger, \id_C\rangle, h^\dagger \rangle,
\end{eqnarray*}
for all $f: A \x B \x C \to A$ and $g: A \x B \x C \to B$, where 
$h = g \circ \langle f^\dagger, \id_{B \x C}\rangle : B \x C \to B$.

This identity was independently found in \cite{Bekic} and \cite{DeBakkerScott}. 
As is well-known, it asserts that a system 
\begin{eqnarray*}
x &=& f(x,y,z)\\
y &=& g(x,y,z)
\end{eqnarray*} 
can be solved by Gaussian elimination by solving the first equation and substituting 
the solution into the second equation to obtain
\begin{eqnarray*}
x &=& f^\dagger(y,z)\\
y &=& g(f^\dagger(y,z),y,z) = h(y,z),
\end{eqnarray*} 
and then by solving the second equation and substituting the solution into the first
to obtain the final result
\begin{eqnarray*}
x &=& f^\dagger(h^\dagger(z),z)\\
y &=& h^\dagger(z). 
\end{eqnarray*} 

In conjunction with the fixed point and parameter identities, the following 
is a special case of the pairing identity:
\begin{eqnarray}
\label{eq-pairingspec}
\langle f , g \circ (\pi^{A \x B}_2 \x \id_C)\rangle^\dagger &=& \langle f^\dagger \circ \langle g^\dagger, \id_C\rangle, 
g^\dagger\rangle,
\end{eqnarray}
where $f: A \x B \x C \to A$ and $g: B \x C \to B$. In the category $\CL$, it asserts that the least solution
of the system of equations 
\begin{eqnarray*}
x &=& f(x,y,z)\\
y &=& g(y,z)
\end{eqnarray*} 
is $x = f^\dagger(g^\dagger(z),z)$ and $y = g^\dagger(z)$.

{\sc Group identities}

Suppose that $G$ is a finite group whose underlying set is $[n]$.
Let $i\cdot j$ denote the multiplication 
of $i,j\in [n]$. The identity associated with $G$ is: 
$$ \langle f \circ (\rho_1 \x \id_B), \ldots, f \circ (\rho_n \x \id_B) \rangle^\dagger
= \Delta_n \circ (f \circ (\Delta_n \x \id_B))^\dagger$$
where $f: A^n \x B \to A$ and for each $i$, $\rho_i$ denotes the 
function $[n] \to [n]$ given by $j \mapsto i\cdot j$
(as well as the associated morphism  $\rho_i^{A,\ldots,A} = \langle \pi^{A^n}_{i \cdot 1 },\ldots,\pi^{A^n}_{i\cdot n}\rangle
: A^n \to A^n$ and $\Delta_n = \Delta^A_n$ is the diagonal morphism $A \to A^n$
defined above.  

This identity can be explained in the following way. Consider the system 
of equations 
\begin{eqnarray}
\notag
x_1 &=& f(x_{1\cdot 1},\ldots,x_{1\cdot n},y)\\
&\vdots &\label{eq-grp1}\\
\notag
x_n &=& f(x_{n\cdot 1},\ldots,x_{n \cdot n},y)
\end{eqnarray} 
and the single equation
\begin{eqnarray}
\label{eq-grp2}
x &=& f(x,\ldots,x,y).
\end{eqnarray}
Then the group identity associated with $G$ asserts that (\ref{eq-grp1}) is equivalent to (\ref{eq-grp2})
in the sense that each component of the least solution of (\ref{eq-grp1}) 
agrees with the least solution of (\ref{eq-grp2}).

Each finite group $G$ (equipped with the natural self action) 
can be seen as a finite automaton, and in a similar fashion, 
one may associate an identity with every finite automaton \cite{Esgroup}. 
These are essentially the commutative identities of \cite{Esikaxioms}. 

\begin{deff}
An iteration category is a cartesian category equipped with a dagger operation 
satisfying either the parameter, fixed point, pairing, permutation and
group (or commutative) identities, or the parameter, composition, 
double dagger and group (or commutative) identities. 
\end{deff} 

The following completeness result is from \cite{Esikaxioms,BEbook}.

\begin{thm}
An identity involving the cartesian category operations and dagger holds in 
$\CL$ with the least fixed point operation as dagger iff it holds in 
all iteration categories. 
\end{thm} 

\begin{remark}
{\rm 
Iteration categories, or iteration theories, were introduced independently 
in \cite{BEW1} and \cite{Esikaxioms}\footnote{In \cite{Esikaxioms}, 
iteration theories were called `generalized iterative theories'.}.
The axiomatization in \cite{Esikaxioms} used the commutative identities. 
It was proved in \cite{Esgroup} that the commutative identities can be 
simplified to the group identities. Moreover, it was shown 
that the identities associated 
with the members of a subclass $\mathcal{G}$ of the finite groups suffices instead 
of all group identities iff every finite group is isomorphic to a quotient of a subgroup
of a group in $\mathcal{G}$, see \cite{Esgroup,Espower}. Nevertheless some further 
simplifications of the axioms are still possible, see \cite{EsAC,EsMSCS2015}. 
}
\end{remark}

We mention one more property that is not an identity, but a quasi-identity. 
It is stronger that the group identities, yet most of the standard models satisfy it.
(Actually the commutative identities were introduced in \cite{Esikaxioms} 
in order to replace this quasi-identity by weaker identities, since
when it comes to equational theories, the best way to present them is by 
providing equational bases.)

{\sc Weak functorial implication}

This axiom asserts that for all $f: A ^n \x B \to A^n$ and $g: A \x B \to  A$,
if  
$f \circ (\Delta_n \x \id_B) = \Delta_n \circ g$, then 
$$f^\dagger = \Delta_n \circ g^\dagger.$$

In $\CL$, this means that if $f = \langle f_1,\ldots,f_n \rangle : A^n \x B \to A^n$ 
and $g : A \x B \to A$ are such that $f_i(x,\ldots,x,y) = g(x,y)$ 
for all $i \in [n]$, then the system of equations 
\begin{eqnarray*}
x_1 &=& f_1(x_1,\ldots,x_n,y)\\
&\vdots & \\
x_n &=& f_n(x_1,\ldots,x_n,y)
\end{eqnarray*} 
is equivalent to the single equation 
\begin{eqnarray*}
x &=& g(x,y).
\end{eqnarray*}

It is clear that if the weak functorial implication holds, then so do the 
group (or commutative) identities. 

\begin{remark}
{\rm 
Sometimes we will apply the least fixed point operation to functions 
$f : A \x B \to A$, where $A,B$ are complete lattices, 
 which are monotonic in the first argument but anti-monotonic 
in the second. Such a function may be viewed as a monotonic 
function $A \x B^d \to A$, where $B^d$ is the dual 
of $B$. Hence, in this case, $f^\dagger$ is a monotonic 
function $B^d \to A$, or --as we will consider it-- an anti-monotonic function 
$B \to A$. More generally, we will also consider functions 
that are monotonic in some arguments and anti-monotonic in 
others, but always take the least fixed point w.r.t. an 
argument in which the function is monotonic. 
}
\end{remark} 

\section{The category $\bCL$}

The objects of $\bCL$ are complete lattices. Suppose that $A,B$ are complete lattices.
A morphism from $A$ to $B$ in $\bCL$, denoted $f: A \circarrow B$, is a $\leq_p$-monotonic function $f: A \x A \to B \x B$, where $A \x A$ and $B \x B$ are the complete bilatices determined by  
$A$ and $B$. Thus, $f = \langle f_1,f_2\rangle$ such that $f_1 : A \x A \to B$ is monotonic 
in its first argument and anti-monotonic in the 
second argument, and  $f_2: A \x A \to B$ is anti-monotonic in its first argument and monotonic in its second argument. 
(Such functions $f$ are called approximations in \cite{Vennekensetal}.)
Composition is ordinary function composition and for each complete lattice $A$,
the identity morphism $\bid_A : A \circarrow A$ is the identity function 
$\id_{A \x A} = \id_A \x \id_A = \langle \pi^{A \x A}_1,\pi^{A \x A}_2\rangle: A \x A \to A \x A$.

The category $\bCL$ has finite products. (Actually it has all products). 
Indeed, a terminal object of $\bCL$ 
is any $1$-element lattice. Suppose that $A_1,\ldots,A_n$ are complete lattices. 
Then consider the direct product $A_1 \x \cdots \x A_n$  as an object of $\bCL$ 
together with the following morphisms $\bpi^{A_1 \x \cdots \x A_n}_i : A_1 \x \cdots \x A_n \circarrow A_i$,
$ i \in [n]$. For each $i$, $\bpi^{A_1 \x \cdots \x A_n}_i$ is the function 
$$A_1 \x \cdots \x A_n \x A_1 \x \cdots \x A_n \to A_i\x A_i$$
defined by 
\begin{eqnarray*}
\bpi^{A_1\x \cdots \x A_n}_i(x_1,\ldots,x_n, x'_1,\ldots,x'_n) &=&  (x_i,x'_i),
\end{eqnarray*}
so that in $\Set$, $\bpi^{A_1\x \cdots \x A_n}_i$ can be written as  
\begin{eqnarray*}
\langle \pi^{A_1\x \cdots \x A_n \x A_1 \x \cdots \x A_n}_i,
\pi^{A_1\x \cdots \x A_n \x A_1 \x \cdots \x A_n}_{n+i}\rangle
&=& \pi^{A_1\x \cdots \x A_n}_i \x \pi^{A_1\x \cdots \x A_n}_i.
\end{eqnarray*}

It is easy to see that the morphisms $\bpi^{A_1\x \cdots \x A_n}_i$, $i \in [n]$, 
determine a product diagram in $\bCL$. To this end, let $f^i = \langle f^i_1, f^i_2 \rangle 
 : C \circarrow A_i$ in $\bCL$, for all $i\in [n]$, so that each 
$f^i$  is a $\leq_p$-monotonic function $C \x C \to A_i \x A_i$.  
Then let $h = \langle h_1,h_2\rangle$, where 
$h_1 = \langle f^1_1,\ldots,f^n_1\rangle$ and $h_2 = \langle f^1_2,\ldots,f^n_2 \rangle$ 
in the category $\CL$. Thus, $h_1$ and $h_2$ are functions $C \x C \to A_1 \x \cdots \x A_n$.

We prove that $h$ is the 
target tupling of $f^1,\ldots,f^n$ in $\bCL$. First, since each $f^i_1$ 
is monotonic in its first argument and anti-monotonic in the second argument, 
the same holds for $h_1$. In the same way, $h_2$ is anti-monotonic
in the first argument and monotonic in the second. 
Thus, $h$ is $\leq_p$-monotonic. 
 Next, writing just 
$\bpi_i$ for $\bpi^{A \x \cdots \x A_n}_i$ and $\pi_i$ for 
$\pi^{A \x \cdots \x A_n}_i$, 
 where $i \in [n]$, we have 
\begin{eqnarray*}
\bpi_i \circ h
&=& 
\bpi_i \circ \langle h_1,h_2 \rangle\\
&=& 
(\pi_i \x \pi_i) \circ \langle \langle f^1_1,\ldots,f^n_1 \rangle ,  \langle f^1_2,\ldots,f^n_2 \rangle \rangle  \\
&=& 
\langle \pi_i \circ  \langle f^1_1,\ldots,f^n_1 \rangle , \pi_i \circ \langle f^1_2,\ldots,f^n_2 \rangle \rangle \\
&=& 
\langle f^i_1,f^i_2\rangle  \\
&=& 
f_i.
\end{eqnarray*}
It is also clear that $h$ is the unique morphism 
$C \circarrow A_1 \x \cdots \x A_n$ in $\bCL$ with this property.

\begin{prop}
$\bCL$ is a cartesian category in which the product of any objects $A_1,\ldots,A_n$ 
agrees with their product in $\CL$. 
\end{prop}

By the above argument, the tupling of any sequence of morphisms 
$f^i = \langle f^i_1,f^i_2 \rangle: C \circarrow A_i$ in $\bCL$ is $h = \langle h_1,h_2\rangle$, 
where $h_1$ is the tupling of the $f^i_1$ and $h_2$ is the tupling of the 
$f^i_2$ in $\Set$. We will denote it by $\blangle f^1,\ldots,f^n\brangle: C \circarrow A_1 \x \cdots \x A_n$.

For further use, we note the following. 
Suppose that $\rho: [m]\to [n]$ and $A_1,\ldots,A_n$ are complete lattices. 
Then the associated morphism  $\brho^{A_1,\ldots,A_n} : A_1\x \cdots \x A_n 
\circarrow A_{\rho(1)} \x \cdots \x A_{\rho(m)}$ 
in $\bCL$ is the function
$$A_1\x \cdots \x A_n \x A_1\x \cdots \x A_n \to  A_{\rho(1)} \x \cdots \x A_{\rho(m)} \x A_{\rho(1)} \x \cdots \x A_{\rho(m)}$$
given by 
$$(x_1,\ldots,x_n,x'_1,\ldots,x'_n)  \mapsto (x_{\rho(1)},\ldots,x_{\rho(m)}, x'_{\rho(1)},\ldots,x'_{\rho(m)}).$$
Thus, 
$$\brho^{A_1,\ldots,A_n} = \rho^{A_1,\ldots,A_n} \x  \rho^{A_1,\ldots,A_n},$$
where $\rho^{A_1,\ldots,A_n}$ is the morphism associated with $\rho$ 
and $A_1,\ldots,A_n$ in $\Set$ (or $\CL$). This is in accordance with 
$\bid_A = \id_A \x \id_A$.

Suppose that $f: C \circarrow A$ and $g: D \circarrow B$ in $\bCL$, so that $f$ is 
a function $C \x C \to A \x A$ and $g$ is a function $D \x D \to B \x B$.   Then 
$f \x g : C \x D \circarrow A \x B$ in the category $\bCL$ is the function 
$$(\id_A \x \langle \pi^{B \x A}_2, \pi^{B \x A}_1   \rangle \x \id_B) \circ
h \circ (\id_C \x \langle \pi^{D \x C} _2, \pi^{D \x C}_1   \rangle \x \id_D): C \x D \x C \x D \to A \x B \x A \x B,$$
where $h$ is $f \x g : C\x C \x D \x D \to A \x A \x B \x B$ in $\Set$. Hence,
$h = \langle h_1,h_2\rangle$ with 
\begin{eqnarray*}
h_1(x,y,x',y') &=& (f_1(x,x'), g_1(y,y'))\\
h_2(x,y,x',y') &=& (f_2(x,x'), g_2(y,y')).
\end{eqnarray*}

\subsection{Some subcategories}

Motivated by \cite{Deneckeretalsurvey,DeneckeretalULT,Vennekensetal}, we define several subcategories of $\bCL$.
Suppose that $A,B$ are complete lattices. Following \cite{Deneckeretalsurvey}, we call 
an ordered pair $(x,x') \in A \x A$  \emph{consistent} if $x \leq x'$. Moreover, we call $f: A \circarrow B$ 
in $\bCL$ consistent if it maps consistent pairs to consistent pairs. 
It is clear that if $f : A \circarrow B$ and $g: B \circarrow C$ in $\bCL$ are consistent, 
then so is $g \circ f: A \circarrow C$, moreover, $\bid_A$ is always consistent. 
Also, for any sequence $A_1,\ldots,A_n$ of complete 
lattices, the projections $\bpi^{A_1 \x \cdots \x A_n}_i: A_1 \x \cdots \x A_n \circarrow A_i$, $i \in [n]$  
are consistent. And when $f_i: C \circarrow A_i$, for all $i \in [n]$, then 
$\blangle f_1,\ldots,f_n \brangle : C \circarrow A_1 \x \cdots \x A_n$  is consistent 
iff each $f_i$ is. Hence, the consistent morphisms in $\bCL$ determine a cartesian 
subcategory of $\bCL$ with the same product diagrams. Let $\CCL$ denote this subcategory. 

We define two subcategories of $\CCL$. The first one, $\ACL$, is the subcategory determined by those 
morphisms $f = \langle f_1,f_2\rangle : A \circarrow B$ in $\bCL$ 
such that $f_1(x,x) \leq  f_2(x,x)$ for all $x \in A$. 
The second, $\AsCL$, is the subcategory determined by those $f :A \circarrow B$
with $f_1(x,x) = f_2(x,x)$. These are again cartesian subcategories
with the same product diagrams.

As noted in \cite{Deneckeretalsurvey}, most applications of approximation 
fixed point theory use \emph{symmetric} functions. 
We introduce the subcategory of $\bCL$ 
having complete lattices as object but only symmetric $\leq_p$-preserving functions as 
morphisms. 

Suppose that $f: A \circarrow B$ in $\bCL$, say $f = \langle f_1,f_2\rangle$, 
We call $f$ symmetric if $f_2(x,x') = f_1(x',x)$, i.e., when 
\begin{eqnarray*}
f_2 &=& f_1 \circ \langle \pi^{A\x A}_2,\pi^{A \x A}_1\rangle: A \x A \to B.
\end{eqnarray*}
We will express this condition in a concise way as $f_2 = f_1^\op$.

It is easy to prove that if $f: A \circarrow B$ and $g: B \circarrow C$ 
are symmetric, then so is $g \circ f$. Moreover, $\bid_A$ is always 
symmetric. Thus, symmetric morphisms determine a subcategory of $\bCL$,
denoted $\sCL$. In fact, $\sCL$ is a subcategory of 
$\AsCL$, since when $f = \langle f_1,f_2\rangle : A \circarrow B$ is symmetric, 
then necessarily $f_1(x,x) = f_2(x,x)$ for all $x \in A$.
Moreover, it is again a cartesian subcategory with the same products. 

Since the first component of a symmetric morphism uniquely determines the 
second component, $\sCL$ can be represented as the category whose objects are 
complete lattices having as morphisms $A \circarrow B$ (where $A$ and $B$ are complete lattices)
those functions $f: A \x A \to B$ which are monotonic in the first and anti-monotonic in the 
second argument. Composition, denoted $\bullet$, is then defined as follows. Given $f: A \circarrow B$ 
and $g: B \circarrow C$, $g \bullet f : A \circarrow C$ is the function 
$$g \circ \langle f,f^\op\rangle : A \x A \to C,$$ so that $h(x,x') = g(f(x,x'),f(x',x))$. 
The identity morphism $A \circarrow A$ is the projection $\pi^{A \x A}_1$.

\section{Fixed points}

In this section, we recall from \cite{Deneckeretalsurvey} the construction
of stable and well-founded fixed points. More precisely, only symmetric functions
were considered in \cite{Deneckeretalsurvey}, but it was remarked that 
the construction also works for non-symmetric functions.

Suppose that $f = \langle f_1, f_2\rangle : A \circarrow A$ in $\bCL$, 
so that $f$ is a $\leq_p$-monotonic function $A \x A \to A \x A$. Then $f_1: A \x A \to A$ 
is monotonic in its first argument and anti-monotonic in its second argument,
and $f_2: A \x A \to A$ is monotonic in its second argument and anti-monotonic 
in its first argument. Define the functions $s_1,s_2: A \to A$ by 
\begin{eqnarray*}
s_1(x') &=& \mu x. f_1(x,x')\\
s_2(x) &=& \mu x'. f_2(x,x')
\end{eqnarray*} 
and let $S(f) :  A \x A \to A \x A$ be the function $S(f)(x,x') = (s_1(x'),s_2(x))$. 
Since $s_1$ and $s_2$ are anti-monotonic, $S(f)$ is a morphism $A \circarrow A$ in 
$\bCL$. We call $S(f)$ the \emph{stable function} for $f$. 
It is known that every fixed point of $S(f)$ is a fixed point 
of $f$, called a \emph{stable fixed point} of $f$. We let $f^\triangle$ 
denote the set of all stable fixed points of $f$. Since $S(f)$ 
is $\leq_p$-monotonic, there is a $\leq_p$-least stable fixed point
$f^\ddag$, called the \emph{well-founded fixed point} of $f$.

The above construction can slightly be extended.
Suppose that $f = \langle f_1,f_2\rangle : A \x B \circarrow  A$ in $\bCL$, 
so that $f$ is a function $A \x B \x A \x B \to A \x A$. 
Then $f_1 : A \x B \x A \x B \to A$ is monotonic in its first and second arguments
and anti-monotonic in the third and fourth arguments, 
while $f_2: A \x B \x A \x B \to A$ is monotonic in the third and fourth arguments
and anti-monotonic in the first and second arguments. 
Now let $s_1,s_2 : A \x B \x B \to A$ be defined by 
\begin{eqnarray*}
s_1(x',y,y') &=& \mu x. f_1(x,y,x',y')\\
s_2(x,y,y') &=& \mu x'. f_2(x,y,x',y').
\end{eqnarray*} 
We have that $s_1$ is monotonic in its second argument and 
anti-monotonic in the first and third arguments, and
$s_2$ is monotonic in the third argument and anti-monotonic 
in the first and second arguments. 
Define $S(f): A\x A \x B \x B \to A \x A$ by
\begin{eqnarray*}
S(f)(x,x',y,y') &=& (s_1(x',y,y'), s_2(x,y,y')). 
\end{eqnarray*} 
Then $S(f)$, as a function $(A \x A) \x (B \x B) \to A \x A$, is 
$\leq_p$-monotonic in both of its arguments. We call $S(f)$ 
the stable function for $f$. (Note that $S(f)$ can be considered as a morphism $L \x L' \to L$
of the category $\CL$, where $L$ and $L'$ are the complete bilattices $A \x A$ 
and $B \x B$ considered as complete lattices ordered by the relation $\leq_p$.)
For each $y,y'\in B$, let $f^\triangle(y,y')$ 
denote the set of solutions of the fixed point equation $(x,x') = S(f)(x,x',y,y')$.
Hence, $f^\triangle$ is a function from $B \x B$ to the power set of $A \x A$,
that we call the stable fixed point function. In particular, for each $y,y'\in B$ there is a $\leq_p$-least 
element of $f^\triangle(y,y')$. We denote it by $f^\ddag(y,y')$. Since $S(f)$ is 
$\leq_p$-monotonic, so is $f^\ddag : B \x B \to A \x A$. Hence $f^\ddag: B \circarrow A$ 
in $\bCL$.

We have thus defined a dagger operation $^\ddag$ on $\bCL$, 
called the (parametric) \emph{well-founded fixed point operation}.  
In the next two sections, we investigate the 
equational properties of this operation.

\begin{remark}
\label{rem-pointwise}
{\rm 
The parametric well-founded fixed point operation $^\ddag$ is just the pointwise extension
of the operation defined on morphisms $A \circarrow A$. Indeed,
when $f: A \x B \circarrow A$ and $(y,y') \in B\x B$,
then let $g:  A \circarrow A$ be given by 
$g(x,x') = f(x,y,x',y')$. Then 
$f^\ddag (y,y') = g^\ddag$ and  $f^\triangle(y,y') = g^\triangle$.
}
\end{remark}

\begin{remark}
\label{rem-symm}
{\rm 
Suppose that $f : \two \circarrow \two$ is given by $f(x,x') = (\neg x', \neg x)$. 
Then $f$ is symmetric but $f^\ddag$ is not, since $f^\ddag = (0,1)$. Hence $\sCL$ is not closed 
w.r.t. the parametric well-founded fixed point operation. Let $g : \two \x \two \circarrow \two$ 
be given by $g(x,y,x',y') = (\neg x', \neg x)$. Then $g$ is a morphism in 
$\ACL$. However, $g^\ddag(y,y') = (0,1)$ for
all $y,y' \in \two$, so that $g^\ddag$ is not a morphism in $\ACL$. 
Hence, $\ACL$ is also not closed under the parametric well-founded fixed point operation.
} 
\end{remark}

\begin{remark}
\label{rem-consistent}
{\rm 
We provide an example showing that when $f: A \x B  \circarrow A$ in $\bCL$ is consistent, $f^\ddag$ may not be consistent. Indeed, let $A = \two$ and $B = T$ (terminal object), and let $f: A \circarrow A$ be given by
$f(x,x') = (1,\neg x \vee x')$. Then $f$ is consistent, since 
$f(0,0) = f(0,1) = f (1,1) = (1,1)$, but $f^\ddag = (1,0)$, so that $f^\ddag$ is not consistent. Since $f$ is in fact in $\AsCL$, this example also shows that 
neither $\ACL$ nor $\AsCL$ is closed with respect to the well founded 
fixed point operation. 

Note that the above $f$ is not symmetric. In fact, if $f: A \circarrow A$ is 
symmetric, then $f^\ddag : T \circarrow A$ is consistent. This follows from 
Remark~\ref{rem-pointwise} and Theorem 23 in \cite{Deneckeretalsurvey}. 
}
\end{remark}

\section{Some valid identities}

In this section we establish the parameter, fixed point, permutation and 
group identities and the special case (\ref{eq-pairingspec}) 
of the pairing identity for the parametrized well-founded fixed 
point operation over $\bCL$. In fact, we prove that the weak functorial 
implication holds.

\begin{prop}
The parameter identity holds: 
\begin{eqnarray*}
(f \circ (\bid_A \x g))^\ddag &=&  f^\ddag \circ g,
\end{eqnarray*}
for all $f: A \x B \circarrow A$ and $g: C \circarrow B$. 
\end{prop}

{\sl Proof.} Let $h = f \circ (\bid_A \x g)  : A\x C \circarrow A$. Then $S(h) : A \x A \x C \x C \to A \x A$ 
is given by
\begin{eqnarray*}
S(h)(x,x',z,z') &=& (\mu x. f_1 (x,g_1(z,z'), x', g_2(z,z')), \mu x'. f_1(x, g_1(z,z'), x', g_2(z,z')))\\
&=& S(f)((\id_{A \x A} \x g)(x,x',z,z')),
\end{eqnarray*} 
where $f = \langle f_1,f_2\rangle$ and $g = \langle g_1,g_2\rangle$.
Thus, $S(h) = S(f)\circ (\id_{A \x A} \x g)$ in $\Set$ (or $\CL$)
and therefore $h^\triangle = f^\triangle \circ (\id_{A \x A} \x g)$.  Moreover, 
$h^\ddag = f^\ddag \circ g$, since the parameter identity holds for the 
least fixed point operation over $\CL$. \eop

\begin{prop}
The fixed point identity holds: 
\begin{eqnarray*}
f\circ \blangle f^\ddag,\id_B \brangle &=& f^\ddag,
\end{eqnarray*}
for all $f : A \x B \circarrow A$.
\end{prop}

{\sl Proof.} 
By Remark~\ref{rem-pointwise}, it is sufficient to prove our claim only in 
the case when $f: A \circarrow A$, i.e., $f$ is a
$\leq_p$-monotonic function $ A \x A \to A \x A$. 
But it is known that if $f: A \circarrow A$, then each stable fixed point of $f$ 
is a ($\leq_t$-minimal) fixed point, so $f \circ f^\ddag = f^\ddag$. 
(We also have $f \circ f^\triangle = f^\triangle$.) \eop

\begin{prop}
The permutation identity holds: 
\begin{eqnarray*}
(\brho \circ f \circ (\brho^{-1} \x \bid_B))^\ddag &=& \brho \circ f^\ddag,
\end{eqnarray*}
for all $f : A_1\x \cdots \x A_n \x B \circarrow A_1\x \cdots \x A_n$
and permutation $\rho: [n]\to [n]$. 
\end{prop} 

{\sl Proof.} 
We prove this only when $B$ is the terminal object, so that 
$f$ can be viewed as a morphism $f = \langle f_1,f_2 \rangle 
 : A_1\x \cdots \x A_n  \circarrow A_1\x \cdots \x A_n$,
where $f_1,f_2$ are appropriate functions 
$$A_1\x \cdots \x A_n \x A_1\x \cdots \x A_n \to A_1\x \cdots \x A_n.$$
Let $g = \brho \circ f \circ \brho^{-1}$ in $\bCL$, so that $g = \langle g_1,g_2\rangle$
where $g_1,g_2$ are functions 
$$A_{\rho(1)}\x \cdots \x A_{\rho(n)} \x A_{\rho(1)} \x \cdots \x A_{\rho(n)} \to 
A_{\rho(1)}\x \cdots \x A_{\rho(n)}.$$
First we show that 
\begin{eqnarray}
\label{eq-perm1}
S(g) &=& \brho \circ S(f) \circ \brho^{-1}
\end{eqnarray}
in $\bCL$, i.e., 
\begin{eqnarray*}
S(g) &=& (\rho \x \rho) \circ S(f) \circ (\rho^{-1} \x \rho^{-1})
\end{eqnarray*}
in $\Set$ (or $\CL$). 
Below we will denote by $x,x'$ $n$-tuples
in $A_1\x \cdots \x A_n$. Similarly, let $y,y'$ denote $n$-tuples in 
$A_{\rho(1)}\x \cdots \x A_{\rho(n)}$.
Note that if $x = (x_1,\ldots,x_n) \in A_1\x \cdots \x A_n$, 
then $\rho(x) = (x_{\rho(1)},\ldots,x_{\rho(n)})$ in  $A_{\rho(1)}\x \cdots \x A_{\rho(n)}$.
And if $y = (y_1,\ldots,y_n) \in A_{\rho(1)}\x \cdots \x A_{\rho(n)}$,
then $\rho^{-1}(y) = (y_{\rho^{-1}(1)},\ldots,y_{\rho^{-1}(n)})$ in $A_1\x \cdots \x A_n$.
Let 
\begin{eqnarray*}
s_1(x') &=& \mu x. f_1(x,x')\\
s_2(x) &=& \mu  x'. f_2(x,x').
\end{eqnarray*} 
Then $S(f)(x,x') = (s_1(x'), s_2(x))$. 
Similarly, let 
\begin{eqnarray*}
t_1(y') 
&=& \mu y. \rho(f_1(\rho^{-1}(y), \rho^{-1}(y')))\\
t_2(y) 
&=& \mu y'. \rho(f_2(\rho^{-1}(y), \rho^{-1}(y'))).
\end{eqnarray*}
Then $S(g)(y,y') = (t_1(y'), t_2(y))$. Since the permutation and parameter identities 
hold for the least fixed point operation over $\CL$, we obtain that 
\begin{eqnarray*}
t_1(y') &=& \rho(s_1(\rho^{-1}(y'))\\
t_2(y) &=& \rho(s_2(\rho^{-1}(y)),
\end{eqnarray*} 
proving (\ref{eq-perm1}). Now from (\ref{eq-perm1}), since the permutation identity holds
for the least fixed point operation over $\CL$, it follows that $g^\ddag = \brho \circ f^\ddag$ in $\bCL$. 
Moreover, it follows that the stable fixed points of $g$ are of the form $(\rho(x), \rho(x'))$, 
where $(x,x')$ is a stable fixed point of $f$. (A suggestive notation: $g^\triangle = \rho \circ f^\triangle$.) \eop

We now establish a special case of the pairing identity. It will be shown later that 
the general form of the identity does not hold. 

\begin{prop}
\label{prop-pairingspec}
The identity (\ref{eq-pairingspec}) holds: 
\begin{eqnarray*}
\blangle f , g \circ (\bpi^{A \x B}_2 \x \bid_C)\brangle^\ddag &=& \blangle f^\ddag \circ \blangle g^\ddag, \bid_C\brangle, 
g^\ddag\brangle,
\end{eqnarray*}
where $f: A \x B \x C \circarrow A$ and $g: B \x C \circarrow B$.
\end{prop}

{\sl Proof.} It suffices to consider the case when there is no parameter. So let 
$f = \langle f_1, f_2\rangle: A \x B \circarrow A$ and $g = \langle g_1,g_2\rangle : B \circarrow B$,
so that $f_1, f_2: A \x B \x A \x B \to A$ and $g_1,g_2: B \x B \to B$. 
Let $ h = \blangle f , g \circ \bpi^{A \x B}_2 \brangle : 
A \x B \circarrow A \x B$ in $\bCL$. 
Then $h^\ddag$ can be constructed as 
follows. First consider 
\begin{eqnarray*}
 &&\mu (x,y). (f_1(x,y,x',y'), g_1(y,y'))\quad {\rm and}\\
 &&\mu (x',y'). (f_2(x,y),x',y'), g_2(y,y')).
\end{eqnarray*}
Since (\ref{eq-pairingspec}) and the parameter identity hold for the least fixed point operation over $\CL$, 
we know that these functions can respectively be written as
\begin{eqnarray*}
 &&(\mu x. f_1(x,\mu y. g_1(y,y'), x',y'), \mu y. g_1(y,y'))\quad {\rm and}\\
 &&(\mu x'. f_2(x, y, x', \mu y'.g_2(y,y')), \mu y'. g_2(y,y')).
\end{eqnarray*}
Now $h^\ddag$ can be obtained 
by solving the system of equations 
\begin{eqnarray*}
(x,x') &=& (\mu x. f_1(x,\mu y. g_1(y,y'), x',y'), \mu x'. f_2(x, y, x', \mu y'. g_2(y,y')) = S(f)((x,x'), S(g)(y,y'))\\
(y,y') &=& (\mu y. g_1(y,y'), \mu y'. g_2(y,y')) = S(g)(y,y')
\end{eqnarray*}
for its least solution w.r.t. $\leq_p$. Moreover, it follows that 
$h^\triangle$ consists of all $((x,y),(x',y'))$ such that $(y,y')$ is a 
stable fixed point of $g$ and $(x,x')$ is in $f^\triangle(y,y')$. 
In particular, since the least fixed point operation over $\CL$ 
satisfies (\ref{eq-pairingspec}), it holds that 
$h^\ddag  = 
\blangle f^\ddag \circ g^\ddag, g^\ddag \brangle$ as claimed. \eop 

\begin{remark} 
{\rm 
The identity (\ref{eq-pairingspec}) has already been established in 
Theorem 3.11 of \cite{Vennekensetal}, see also the  
Splitting Set Theorem of \cite{LifschitzTurner}.
}
\end{remark}

\begin{prop}
The weak functorial dagger implication holds: for all $f: A ^n \x B \circarrow A^n$ and $g: A \x B \circarrow A$ in 
$\bCL$: if $f \circ (\bDelta_n \x \bid_B) = \bDelta_n \circ g$, then $f^\ddag = \bDelta_n \circ g^\ddag$. 
\end{prop}

{\sl Proof.} We spell out the proof only in the case when $B$ is a terminal object. 
So let $f : A^n \circarrow A^n$ and $g: A \circarrow A$ 
in $\bCL$, say $f = \langle f_1,f_2\rangle$ and $g = \langle g_1,g_2\rangle$, where 
$f_i: A^n\x A^n \to A^n$ and $g_i: A\x A \to A$ are appropriate functions for $i = 1,2$.

The assumption $f \circ \bDelta_n  = \bDelta_n \circ g$ can be rephrased 
as $$f_i \circ ( \Delta_n \x \Delta_n) = \Delta_n \circ g_i,\quad i= 1,2,$$
i.e.,
\begin{eqnarray*} 
f_1(x,\ldots, x,x', \ldots,x') &=&  (g_1(x,x'),\ldots,g_1(x,x'))\\
f_2(x,\ldots, x,x', \ldots,x') &=&  (g_2(x,x'),\ldots,g_2(x,x'))
\end{eqnarray*} 
for all $x,x' \in A$. 
Since the weak functorial dagger implication and the parameter identity 
hold for the least fixed point operation over $\CL$, it follows that 
\begin{eqnarray*}
h_1(x',\ldots,x') &=& (k_1(x'),\ldots,k_1(x'))\\
h_2(x,\ldots,x) &=& (k_2(x),\ldots,k_2(x))
\end{eqnarray*} 
where $h_1(x_1',\ldots,x_n')$ and $h_2(x_1,\ldots,x_n)$ 
are respectively the least solutions of 
\begin{eqnarray*}
(x_1,\ldots,x_n) &=& f_1(x_1,\ldots,x_n,x'_1,\ldots,x'_n)\quad {\rm and}\\
(x_1',\ldots,x_n') &=& f_2(x_1,\ldots,x_n,x'_1,\ldots,x'_n)
\end{eqnarray*} 
and $k_1(x')$ and $k_2(x)$ denote the least solutions of 
\begin{eqnarray*}
x & =& g_1(x,x')\quad{\rm and}\\
x' &=& g_2(x,x'),
\end{eqnarray*} 
so that $S(f)(x_1,\ldots,x_n,x_1',\ldots,x_n') = (h_1(x_1',\ldots,x_n'), h_2(x_1,\ldots,x_n))$,
moreover, 
$S(g)(x,x') = (k_1(x'), k_2(x))$. 
Consider now the equations 
\begin{eqnarray*}
(x_1,\ldots,x_n,x'_1,\ldots,x'_n) &=& (h_1(x_1',\ldots,x_n'), h_2(x_1,\ldots,x_n))
\end{eqnarray*} 
and 
\begin{eqnarray*}
(x,x') &=& (k_1(x'),k_2(x)).
\end{eqnarray*} 
Since the weak functorial dagger implication and the parameter identity 
hold for the least fixed point operation over $\CL$, the 
$\leq_p$-least solution of the first equation can be obtained as the 
$2n$-tuple whose first $n$ components are equal to the first component 
of the $\leq_p$-least solution of the second equation, and whose second $n$ 
components are equal to the second component of the $\leq_p$-least solution of the second equation. 
This means that $f^\ddag = (\Delta_n \x \Delta_n) \circ g^\ddag$ in $\Set$, i.e., 
$f^\ddag = \bDelta_n \circ g^\ddag$ in $\bCL$. (It also holds that if $(x,x')$ 
is a stable fixed point of $g$, then $(x,\ldots,x,x',\ldots,x')$ is a stable fixed 
point of $f$.) \eop 

\begin{cor}
The identities associated with finite groups hold for the 
parametrized well-founded fixed point operator over $\bCL$. 
\end{cor} 

In fact, each identity associated with a finite automaton holds.

\section{Some identities that fail}

\begin{prop}
The composition identity fails even in the following simple case:
\begin{eqnarray*}
f \circ (f \circ f)^\ddag &=& ( f \circ f)^\ddag,
\end{eqnarray*}
where $f : A \circarrow A$. 
\end{prop}

{\sl Proof.} 
Let $f : \two \circarrow \two$ be given by $f(x,x') = (\neg x', \neg x)$ (see also Remark~\ref{rem-symm}). 
Then $f \circ f$ is the identity function on $\two \x \two$,
hence $(f \circ f)^\ddag = (0,0)$.   On the other hand, 
$f \circ (f \circ f)^\ddag = (1,1)$. \eop

\begin{prop}
The squaring identity $(f \circ f)^\ddag = f^\ddag$ fails, where $f: A \circarrow A$. 
\end{prop}

{\sl Proof.} Let $f$ be as in the previous proof. Then $(f \circ f)^\ddag = (0,0)$ 
as shown above. But $f^\ddag = (0,1)$. \eop

Since the fixed point, parameter and permutation identities hold
but the composition identity fails, the pairing identity also must fail,
see \cite{BEbook}. 
We can give a direct proof.

\begin{prop}
The pairing identity 
\begin{eqnarray*}
\langle f,g\rangle^\ddag &=& \langle f^\ddag \circ \langle h^\ddag, \bid_C\rangle, h^\ddag\rangle,
\end{eqnarray*}
where $h = g \circ \blangle f^\ddag, \bid_{B \x C}\brangle$ fails, where $f: A \x B \x C \circarrow A$ 
and $g: A \x B \x C \circarrow B$. 
\end{prop} 

{\sl Proof.} 
Let $f,g : \two \x \two \circarrow \two$ in $\bCL$, 
so that $f$ and $g$ are appropriate functions 
$\two \x \two \x  \two \x \two \to \two \x \two$,
\begin{eqnarray*}
f(x,y,x',y') &=& (\neg y', \neg y)\\
g(x,y,x',y') &=& (\neg x',  \neg x ).
\end{eqnarray*} 
Then 
\begin{eqnarray*}
\blangle f,g \brangle(x,y,x',y') &=& (\neg y',\neg x',\neg y,\neg x)
\end{eqnarray*}
and thus $\blangle f,g \brangle^\ddag = (0,0,1,1)$.
On the other hand, $f^\ddag(y,y') = (\neg y', \neg y)$, 
hence $h = g \circ \blangle f^\ddag, \bid_\two \brangle$
is the identity function on $\two \x \two$ and $h^\ddag = (0,0)$
and $f^\ddag \circ h^\ddag = (1,1)$. It follows that 
$\blangle f^\ddag \circ h^\ddag , h^\ddag\brangle = (1,0,1,0)$.  
\eop 

Each of the above examples involved symmetric morphisms. We now 
refute the double dagger identity, but we use a non-symmetric
morphism. 

\begin{prop}
The double dagger identity fails in $\bCL$.
\end{prop}

{\sl Proof.} Let $g: \two \x \two \circarrow \two$ be given by
$g(x,y,x',y') = (\neg y', \neg x)$, and let 
$h = g \circ \blangle \bid_\two,\bid_\two \brangle : \two \circarrow \two$,
so that $h(x,x') = (\neg x',\neg x)$. We already know that 
$h^\ddag = (0,1)$. But $g^\ddag(y,y') = (\neg y',y)$ and
$g^{\ddag\ddag} = (1,0)$. \eop 

\section{Conclusion}

We extended the well-founded fixed point operation of 
\cite{Deneckeretalsurvey,Vennekensetal} to a parametric operation and studied
its equational properties. We found that several of the 
identities of iteration theories hold for the parametric 
well-founded fixed point operation, but some others fail. 
Two interesting questions for further investigation arise.
The first one concerns the \emph{algorithmic description} of the valid 
identities of the well-founded fixed point operation. 
Does there exist an algorithm to decide whether an identity
(in the language of cartesian categories equipped with a
dagger operation) holds for the well-founded fixed point
operation? The second one concerns the \emph{axiomatic description} of 
the valid identities of the well-founded fixed point 
operation. These questions are relevant in connection with modular
logic programing, cf. \cite{Ferrarisetal,Janhunenetal,LifschitzTurner}. 

An alternative semantics of logic programs with negation based on 
an infinite domain of truth values was proposed
in \cite{RondogiannisWadge}. The infinite valued  approach has been 
further developed in the abstract setting of `stratified complete lattices' in 
\cite{Charalambidisetal,EsikRondogiannis2,EsikRondogiannis1,EsikWOLLIC,EsikTbiLLC}. 
In particular, it has been proved in \cite{EsikWOLLIC} that the 
stratified least fixed point operation arising in this approach
does satisfy all identities of iteration theories. 

{\bf Acknowledgments} The authors would like to thank Panos Rondogiannis for pointing 
out some of the references. The second author would like to thank the hospitality of the 
Institute of Informatics Gaspard Monge of Universt\'e Paris Est.

\end{document}

%% file: four.tex
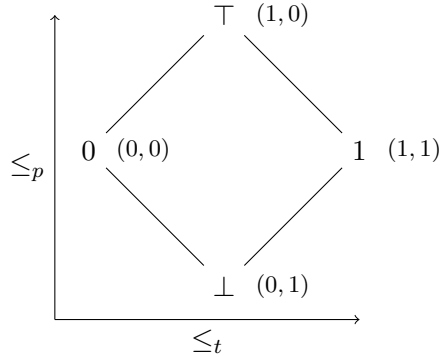
\begin{figure}[htb]
\begin{center}
\begin{tikzpicture}[scale=1.8, auto,swap]
	\node[label=right:{\footnotesize $(0,1)$}] (bot) at (0,0) {$\bot$};
	\node[label=right:{\footnotesize $(1,0)$}] (top) at (0,2) {$\top$};
	\node[label=right:{\footnotesize $(0,0)$}] (false) at (-1,1) {$0$};
	\node[label=right:{\footnotesize $(1,1)$}] (true) at (1,1) {$1$};
	
	\path[draw] (bot) -- (true);
	\path[draw] (bot) -- (false);
	\path[draw] (true) -- (top);
	\path[draw] (false) -- (top);
	
	\path[draw,->] (-1.25,-0.25) to node[below] {$\leq_t$} (1,-0.25);
	\path[draw,->] (-1.25,-0.25) to node[left] {$\leq_p$} (-1.25,2);
\end{tikzpicture}
\end{center}
\caption{A representation of $\mathbf{4} \approx \mathbf{2} \times \mathbf{2}$ taken from \cite{Fittingsurvey}.\label{figure:four}}
\end{figure}